\documentclass[prb,aps,twocolumn,showpacs,bibnotes,superscriptaddress,epsf]{revtex4-1}

\usepackage{graphicx}
\usepackage{amsmath}
\usepackage{amssymb}
\usepackage{color}
\usepackage{dcolumn}
\usepackage{epsfig}
\usepackage{bm}
\usepackage{array}
\usepackage{multirow}
\usepackage[urlcolor=blue]{hyperref}
\usepackage{booktabs}
\usepackage{float}
\hypersetup{colorlinks=true, linkcolor=blue, citecolor=blue}

\begin{document}

\title{Universal phonon softening in the pseudogap state of Tl$_2$Ba$_2$Ca$_{n-1}$Cu$_n$O$_{2n+4+\delta}$  }

\author{Jia-Wei Hu}
\affiliation{Key Laboratory of Materials Physics, Institute of Solid State Physics, HFIPS, Chinese Academy of Sciences, Hefei 230031, China}
\affiliation{University of Science and Technology of China, Hefei 230026, China}
\affiliation{Center for High Pressure Science and Technology Advanced Research, Shanghai 201203, China}

\author{Kai Zhang}
\affiliation{Center for High Pressure Science and Technology Advanced Research, Shanghai 201203, China}

\author{Yong-Chang Ma}
\affiliation{School of Materials Science and Engineering, Tianjin University of Technology, Tianjin 300384, China}

\author{Nan-Lin Wang}
\affiliation{International Center for Quantum Materials, School of Physics, Peking University, Beijing 100871, China}

\author{Viktor V. Struzhkin}
\affiliation{Center for High Pressure Science and Technology Advanced Research, Shanghai 201203, China}

\author{Alexander F. Goncharov}
\affiliation{Earth and Planets Laboratory, Carnegie Institution for Science, Washington, DC 20015, USA}

\author{Hai-Qing Lin}
\affiliation{School of Physics, Zhejiang University, Hangzhou 310058, China}

\author{Xiao-Jia Chen}
\email{xjchen2@gmail.com}
\affiliation{Center for High Pressure Science and Technology Advanced Research, Shanghai 201203, China}
\affiliation{School of Science, Harbin Institute of Technology, Shenzhen 518055, China}
\date{\today}

\begin{abstract}

Exploring the origin of the pseudogap is important for the understanding of superconductivity in cuprates.  Here we report a systematical experimental study on the phonon vibrational properties of Tl$_2$Ba$_2$Ca$_{n-1}$Cu$_n$O$_{2n+4+\delta}$ ($\emph{n}$$=$1,2,3) single crystals based on the Raman scattering measurements over the temperature range from 10 to 300 K.  The temperature evolution of the frequency and linewidth of the observed phonon modes in each member of this family does not follow the expected self-energy effect when entering the superconducting state. Instead, these phonon modes exhibit a universal softening behavior below the temperature around 150 K, which is higher above the superconducting transition. From the comparison with the existing experimental data for various orders, we find that the observed starting temperature for the phonon softening corresponds to the onset opening temperature of the pseudogap. This finding indicates a large lattice effect in the pseudogap state and the non-negligible spin-phonon coupling for such a phonon softening.

\end{abstract}

\maketitle

\section{INTRODUCTION}

The pseudogap has been widely investigated in cuprate superconductors\cite{Timusk,norman}. Its origin and connection to the superconducting gap are two key issues in the understanding of the  mechanism of the superconductivity in cuprates. The spectroscopic evidence was given for a pseudogap in the normal state of underdoped  Bi$_2$Sr$_{2}$CaCu$_2$O$_{8+\delta}$ by using angle-resolved photoemission spectroscopy (ARPES) \cite{Ding1}. The pseudogap was found to have the similar $d$-wave symmetry as the superconducting gap in the underdoped regime\cite{Ding1}. Similar observations were reported in single layer Bi$_2$Sr$_{2-x}$La$_x$CuO$_{6+\delta}$$\cite{Harris}$. Further work on underdoped La$_{2-x}$Sr$_x$CuO$_4$ showed that the opening of the pseudogap and the superconducting gap occur with almost the same energy scale\cite{Xu}. All these phenomena suggested a precursor pairing of the electrons in the pseudogap state$\cite{Ding, Meng, Kanigel}$. Notably, the evidence for decoupling of the pseudogap state from the superconducting state was also reported\cite{Hashimoto2, Yang}. For example, the pseudogap state was found along with a particle-hole asymmetry in Pb$_{0.55}$Bi$_{1.5}$Sr$_{1.6}$La$_{0.4}$CuO$_{6+\delta}$ and Bi$_2$Sr$_2$CaCu$_2$O$_{8+\delta}$\cite{Hashimoto2, Yang}, which was distinct from superconducting state. Recently, the spin textures in the pseudogap state of YBa$_{2}$Cu$_{3}$O$_{6+x}$ have been observed at the nanometer scale\cite{zwang}, supporting the magnetic order characterizing the pseudogap phase in cuprates reported previously\cite{fauq,yli}. Meanwhile, ARPES measurements revealed that the pseudogap exhibits an antagonistic behavior near the superconducting transition temperature $\emph{T}$$_c$\cite{Hashimoto}. This behavior classifies the pseudogap as a competing order similar to other competing phases\cite{Ghiringhelli, Silva, Chang}. In fact, the time-reversal symmetry has been observed to break spontaneously just below the opening temperature of the pseudogap\cite{kamin}, indicating a phase transition takes place there.

So far, no consensus has been reached for explaining such complex competitive or cooperative behaviors. One of the well-known scenarios is the spin-charge separation based on the resonating-valence-bond model$\cite{Anderson}$. Within this framework, the pseudogap state is assumed to be a spin-liquid state without the long-range order. The spinons pair, which originates from the separation between the spin and charge, forms a spin-excitation gap (pseudogap)\cite{Timusk, Kordyuk}. Extensive experiments based on Raman scattering spectroscopy have revealed that the two-magnon peak energy evolves with doping in a similar mannor as the pseudogap and pair peak in almost all cuprate superconductors\cite{sugai2,Sugaimagnon,yli2}. These similarities indicate that the high-energy magnetic fluctuations are directly involved in the formation of the pseudogap and the Cooper pairing interaction. Such a coupling from the high-energy excitation to superconducting quasiparticles has been captured from coherent charge fluctuation spectroscopy\cite{mans}. Numerical computation clearly reproduced the two-magnon peak and its gradual development to a quasiparticle response with doping\cite{nlin}. Although the magnetic signature for the pseudogap as well as the magnetic origin for the superconductivity have been indicated from experiments and theory\cite{zwang,fauq,yli,sugai2,Sugaimagnon,yli2,mans,nlin}, some non-negligible lattice effect on the pseudogap state has also been reported\cite{Temprano, Bendele, Lanzara1}.  For instance, upon oxygen isotope substitution ($^{16}$O/$^{18}$O), the onset temperature of the pseudogap remarkably shifts in La$_{2-x}$Sr$_{x}$CuO$_{4}$\cite{Bendele} and HoBa$_{2}$Cu$_{4}$O$_{8}$\cite{Temprano}. The huge oxygen isotope effects indicate a strong electron-phonon interaction for the pseudogap formation. 

Tl$_2$Ba$_2$Ca$_{n-1}$Cu$_n$O$_{2n+4+\delta}$ [abbreviated as Tl-22($n$-1)$n$] ($\emph{n}$$=$1,2,3) superconductors have attracted significant  interest because of their high $T_{c}^{\prime}$s ($>$90 K) at ambient pressure$\cite{sheng,hazen}$. Such high $\emph{T}_c^{\prime}$s allow the study of the physical properties in the superconducting and normal state over a wide temperature range. The investigations on the homogeneous layered family have been proven to be powerful and effective to draw important information such as lattice effect on superconductivity in cuprates$\cite{Chen1, Chen2, Chen3}$. The crystal structures of Tl-based superconductors have common features with the other high-$\emph{T}$$_c$ superconductors$\cite{Tsvetkov, Liu2, Kovaleva, Cox, Yamauchi}$. Much effort has been devoted to study the normal and superconducting state of Tl-based superconductors$\cite{Kambe, Gerashenko, Zheng, Zetterer, Chrzanowski, Mukherjee, Gasparov, Matsuishi}$. In particular, Raman scattering at low-temperatures can probe both phonon and electronic states as well as their interaction near the Fermi surface$\cite{Thomsen1,Burns,Boekholt,Loa,Zhou,mfli,Hewitt,xjc}$. The vibrational modes are sensitive to the structure and symmetry change. The possible concerns such as defects and the impurity effects can be safely removed by using high-quality single crystals. Therefore, the Raman scattering spectroscopy has been widely used to study many important physical properties of cuprates such as the superconducting gap$\cite{frie,yama,neme,xkchen,Kendziora,rubh,hewi,masui}$, energy scales\cite{sugai2,Sugaimagnon,yli2,tacon}, Cooper-pair density\cite{blanc}, charge density wave$\cite{Loret1, Loret2}$, pseudogap$\cite{Loret3, LimonovP,tmasui,blumb}$, and superconducting fluctuations\cite{blumb}, etc. Currently, the Raman measurements concentrated on the physics in the pseudogap regime of Tl-based family are still missing but desired.

To address all the above mentioned issues, we perform systematic Raman scattering measurements on Tl$_2$Ba$_2$Ca$_{n-1}$Cu$_n$O$_{2n+4+\delta}$ ($\emph{n}$$=$1,2,3) single crystals at the interesting temperature range. The collected phonon modes are carefully analyzed by using the Fano profile. The obtained frequency and linewidth of each phonon mode allow to establish a connection among various orders. Their temperature-dependent behaviors are discussed and the possible origin of the pseudogap state is given based on the observations.

\section{EXPERIMENTAL DETAILS}

The single phase Tl-based crystals synthesized by flux method were detailed previously$\cite{Ma}$. Tl-2201, Tl-2212, and Tl-2223 crystals have  body-centered tetragonal lattice with $I$4/$mmm$ space group. Furthermore, no other phases could be detected from the diffraction patterns. Each sample has a good $c$-axial orientation which is perpendicular to the sample's natural growth surface. The magnetic susceptibility measurements were used to determine the superconducting transition for the Tl-2201, Tl-2212, and Tl-2223 sample at temperature of 92, 109, and 119 K, respectively\cite{Ma,jbz}.

The parallelepiped-shaped, as-grown samples have been cleaved into a size of 100$\times$100$\times$20 $\mu$m $^{3}$ with a smooth surface and their $c$ axis along the thinnest dimension. All the prepared samples were fixed with liquid adhesive in a cryogenic vacuum chamber. Raman spectra were measured with the z(x$^{\prime}$y$^{\prime}$)\={z} polarization in a near-backscattering geometry. A 488 nm laser was used to excite the Raman response. The laser power was kept at 1.5 mW to reduce the heating effect. An 1800 g/mm grating was used to obtain the spectra. The scattered light was recorded with a charge-coupled device designed by Princeton Instruments (PIX100BRX-SF-Q-F-A). A continuous flow cryostat with a high vacuum (6.1$\times$10$^{-7}$ mbar) was used for controlling the sample temperature from 10 K to room temperature .

\section{RESULTS}

\begin{figure}[tbp]
\centering
\includegraphics[width=\columnwidth]{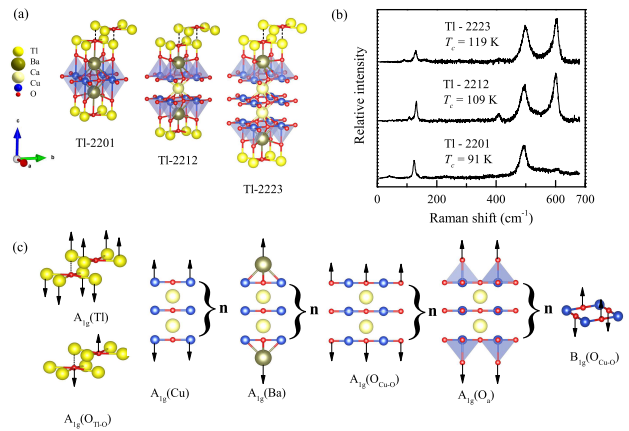}
\caption{(a) Tetragonal structure ($I$4/$mmm$) of Tl$_2$Ba$_2$Ca$_{n-1}$Cu$_n$O$_{2n+4+\delta}$ ($\emph{n}$$=$1,2,3). (b) The representative Raman spectra of Tl-2201, Tl-2212 and Tl-2223 collected with the 488 nm wavelength excitation at room temperature. The $\emph{T}$$_c$ value is taken from the susceptibility measurements. (c) The symmetry of the phonon vibrational modes observed experimentally in Tl-based system. The arrows show the direction of the vibrational modes.}
\end{figure}

\subsection{Assignment of the Raman modes}

The crystal structure of Tl$_2$Ba$_2$Ca$_{n-1}$Cu$_n$O$_{2n+4+\delta}$ ($\emph{n}$$=$1,2,3) is shown in Fig. 1(a). To classify the lattice vibrations and combine them with the theoretical shell model, we consider the space group $I$4/$mmm$ as the symmetry in layered copper oxides. The structures differ from each other by the number of consecutive CuO$_{2}$ layers. The corner sharing square-planar CuO$_{4}$ group is oriented parallel to the $c$ direction. The individual CuO$_{2}$ layer is separated by the Ca atom. There are also two rocksalt type Tl-O layers in this structure giving a layer repeating sequence of [TI-Ba-Cu-Ca-Cu-Ba-Tl]$_{n}$ along the $c$ direction.

Figure 1(b) shows a complete set of Raman spectra of Tl-based compounds measured using the 488 nm laser line at room temperature. All Tl-based crystals have only atoms at positions with site symmetry D$_{4h}$, D$_{2h}$, C$_{4v}$, and C$_{2v}$. Strong peaks show up in the z(x$^{\prime}$y$^{\prime}$)\={z} polarization. The spectra contain phonons with the $\emph{A$_{1g}$}$ + $\emph{B$_{1g}$}$ symmetry. The prominent features are located at 124, 162, 491 and 603 cm$^{-1}$ for Tl-2201, at 107, 133, 161, 410, 494 and 602 cm$^{-1}$ for Tl-2212, and at 90, 132, 161, 269, 403, 501 and 608 cm$^{-1}$ for Tl-2223. Table I gives a  list  of the Raman-active irreducible modes and the phonon mode assignment for Tl$_2$Ba$_2$Ca$_{n-1}$Cu$_n$O$_{2n+4+\delta}$. The obtained results are generally consistent with the earlier studies$\cite{Timofeev, Kulkarni1, Kulkarni2, McCarty, McCarty1}$.

In Fig. 1(c), we show the suggested eigenvector patterns corresponding to the eigenmodes of $\emph{A$_{1g}$}$ symmetry and $\emph{B$_{1g}$}$ symmetry for the Tl-based compounds. However, the $\emph{B$_{1g}$}$ mode of the oxygen ions in the Cu-O planes is hard to be detected in the measurements. Early Raman spectroscopy studies on Tl-based samples also reported the absence of such $\emph{B$_{1g}$}$ phonon branches$\cite{Gasparov,Timofeev, Kulkarni1, Kulkarni2, McCarty, McCarty1}$. The relatively low Raman activity of the $B_{1g}$ modes in Tl-based system was interpreted due to the relatively small perturbations\cite{Gasparov} in their more uniform structures$\cite{Cox}$.

\begin{table*}[tbp]\footnotesize%
\renewcommand\arraystretch{1.6}
\caption{The irreducible representations of the Raman active modes and phonon assignments of Tl$_2$Ba$_2$Ca$_{n-1}$Cu$_n$O$_{2n+4+\delta}$ ($\emph{n}$$=$1,2,3). The frequencies are expressed in cm$^{-1}$. The literature data are from the works of Kulkarn $et$ $al.$$\cite{Kulkarni1, Kulkarni2}$ and McCarty $et$ $al.$$\cite{McCarty, McCarty1}$ as listed in parentheses.}
\resizebox{\textwidth}{38mm}{
\begin{tabular}{cc c c c c c c c c c c c cccccccc}
\hline\hline
 & & & & & & & & \\
Sample & Raman-active irreducible & $\emph{A$_{1g}$}$(Ba) & $\emph{A$_{1g}$}$(Tl) & $\emph{A$_{1g}$}$(Cu) & $\emph{B$_{1g}$}$(O$_{Cu-O}$) & $\emph{A$_{1g}$}$(O$_{Cu-O}$) & $\emph{A$_{1g}$}$(O$_{a}$) & $\emph{A$_{1g}$}$(O$_{Tl-O}$) \\
 & & & & & & & & \\
\hline
 & & 124 &162 & - & - & - & 491 & 603 \\
$Tl-2201$ & 4$\emph{A$_{1g}$}$ + 4$\emph{E$_{g}$}$ & & & & & & \\
 & & (122) & (167) & - & - & - & (495) & (612) \\
\hline
 & & 107 & 133 & 161 & - & 410 & 494 & 602 \\
$Tl-2212$ & 6$\emph{A$_{1g}$}$ + 7$\emph{E$_{g}$}$ + $\emph{B$_{1g}$}$ & & & & & & \\
 & & (108) & (130) & (158) & - & (407) & (494) & (599) \\
\hline
 & & 90 & 132 & 161 & 269 & 403 & 501 & 608 \\
$Tl-2223$ & 7$\emph{A$_{1g}$}$ + 8$\emph{E$_{g}$}$ + $\emph{B$_{1g}$}$ & & & & & & \\
 & & (99) & (133) & (159) & (275) & (405) & (498) & (601) \\
\hline\hline
\end{tabular}}
\end{table*}

\subsection{Temperature dependence of the Raman spectra}

Figures 2(a)-(c) show the spectrum mappings in the entire studied temperature range, where the red color represents the high scattering intensity and the blue color represents the low intensity. It can be clearly seen that there is a significant redshift for the $\emph{A$_{1g}$}$(Tl), $\emph{A$_{1g}$}$(O$_{a}$) and $\emph{A$_{1g}$}$(O$_{Tl-O}$) mode in the low-temperature region. As shown in Figs. 2(d)-(f), the line shape of the phonon modes at representative temperatures exhibits an asymmetric Fano resonance behavior. Quantitative analysis of the linewidth and frequency of these modes was carried out by fitting the line shapes to the Fano profile$\cite{Hewitt}$ with a linear background,
\begin{equation}
I (\omega) = I_{0}\frac{ (q + \epsilon)^{2}}{1 + \epsilon^{2}} + background~~~,
\end{equation}
where
\begin{equation}
\epsilon = \frac{ (\omega - \omega _{0})}{\gamma}~~~.
\end{equation}
Here, $I_0$ is the relative intensity of the phonon peak, $\omega$ and $\omega$$_0$ are the experimental and resonance frequency, 2$\gamma$ is the full width at the half maximum, and $q$ is the asymmetric factor of the phonon shape.

Typical spectra of the $\emph{A$_{1g}$}$(O$_{a}$) mode in Tl-2201 between 10 and 300 K are shown in Fig. 3(a). The peaks display asymmetric line shapes which suggest an interaction between these discrete phonon states and a broad electronic continuum. The values of $\omega$, $\gamma$, and $q$ as a function of temperature are given in Fig. 3(b). At temperatures above 150 K, an overall $\omega$ softening, a linewidth narrowing, and a small increase of $q$ can be seen around $\emph{T}$$_c$ (red dot lines) when going from the normal to superconducting state. The $\omega$ value shifts downward by 2 cm$^{-1}$ in the temperature range between 150 and 92 K. The accompanying linewidth narrowing in the same temperature range is about 1 cm$^{-1}$, just above the instrument detection limit. The line-shape parameter $q$ fluctuates below $\emph{T}$$_c$. The peak position of the $\emph{A$_{1g}$}$ mode exhibits an obvious phonon softening far above $\emph{T}$$_c$. However, the $\gamma$ remains almost unchanged in the temperature range of 10$-$150 K.

\begin{figure}[tbp]
\includegraphics[width=\columnwidth]{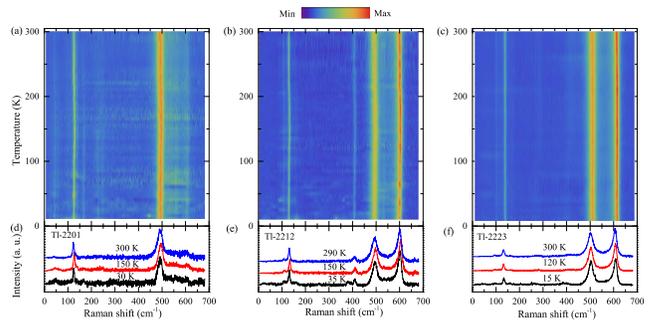}
\caption{ Temperature-dependent Raman active modes. (a)-(c) The mapping graph for Tl-2201, Tl-2212 and Tl-2223, where the red color indicates the high intensity, while the blue color indicates the low intensity. The data points are normalized with respect to the peak value to show the trends of the frequency and width of the phonon peak. (e)-(f) Typical Raman spectra of Tl$_2$Ba$_2$Ca$_{n-1}$Cu$_n$O$_{2n+4+\delta}$ ($\emph{n}$$=$1,2,3) excited by a 488 nm laser at selected temperatures .}
\end{figure}

\subsection{Phonon mode analysis}

The behaviors of $\omega$ and $\gamma$ for Tl$_2$Ba$_2$Ca$_{n-1}$Cu$_n$O$_{2n+4+\delta}$ ($\emph{n}$$=$1,2,3) as a function of temperature are shown in Figs. 4-6, respectively. It is found that the $\emph{A$_{1g}$}$(Tl) mode in each member of this family shifts to higher frequencies on cooling from temperature of 300 to 150 K, followed by a shift back to lower frequencies on cooling through $\emph{T}$$_c$ (Fig. 4). The peak position increases from 162 (at 300 K) to 168 cm$^{-1}$ (at 150 K) for Tl-2201, 131 (at 300 K) to 135 cm$^{-1}$ (at 150 K) for Tl-2212, and 132 (at 300 K) to 133 cm$^{-1}$ (at 150 K) for Tl-2223, respectively. The accompanying $\gamma$ narrowing during the whole temperature range is nearly 1$-$2 cm$^{-1}$. All three curves for the peak position of the $A_{1g}$(Tl) mode are similar demonstrating a maximum above $T_c$, softening on cooling down to $T_c$ and then nearly constant in superconducting state. The data agree with the early report for Tl-2201, Tl-2212 and Tl-2223$\cite{Matsuishi}$ but disagree with those for the single-layer and bilayer samples$\cite{Gasparov}$. Comparing with the Y-based and Bi-based cuprate superconductors$\cite{LimonovP, Loa}$, the low-frequency $\emph{A$_{1g}$}$ modes in the Tl-based system exhibit a mode-hardening behavior at low temperatures.

The temperature dependence of the apical oxygen $\emph{A$_{1g}$}$(O$_{a}$) mode of the Tl-based compounds observed at around 500 cm$^{-1}$ is shown in Fig. 5. Similar to the $\emph{A$_{1g}$}$(Tl) mode, all the frequencies show a monotonic increase over 150 K. For Tl-2201 and Tl-2212, $\omega$ downshifts by 3$-$4 cm$^{-1}$ below $\emph{T}$$_c$, and Tl-2223 exibits softening by about 1 cm$^{-1}$ in the superconducting state. Meanwhile, the $\gamma$ value is reduced by 2.5 cm$^{-1}$ for Tl-2201, 3 cm$^{-1}$ for Tl-2212, and 4 cm$^{-1}$ for Tl-2223 from 300 K to 10 K. Phonon anomalies were observed in the previous Raman studies for the apical oxygen modes$\cite{Chrzanowski, Mukherjee, Matsuishi}$. The $\emph{A$_{1g}$}$(O$_{a}$) mode for the studied Tl-2201, Tl-2212, and Tl-2223 shows a softening behavior at temperature below 150 K but above $T_c$ of each member of this layered family. For comparison, the phonon softening was also found at temperature below $T_c$ in HgBa$_2$Ca$_2$Cu$_3$O$_{8+\delta}$ and Bi$_2$Sr$_2$CaCu$_2$O$_{8+\delta}$$\cite{Zhou, Boekholt, Burns}$. It should be noticed that Tl-based samples exhibit a frequency softening for the apical oxygen model at high temperature above the $T_c$ value. 

The $\emph{A$_{1g}$}$(O$_{Tl-O}$) mode in the charge reservoir layer of the Tl-based materials shows softening again in the frequency across 150 K (Fig. 6). Clearly, phonon softening is about 8 cm$^{-1}$ for Tl-2201, 2 cm$^{-1}$ for Tl-2212, and 1 cm$^{-1}$ for Tl-2223 from 150 K to $\emph{T}$$_c$. The $\gamma$ value has 1 cm$^{-1}$ narrowing from 300 K to 200 K in Tl-2212 and Tl-2223, and becomes constant at the low temperature. The $\gamma$ value nearly does not change in Tl-2201 if there is any variation for the data at lower temperatures ($<$ $\emph{T}$$_c$). The phonon softening was also reported on Tl-based compounds in the early study$\cite{Matsuishi}$. The $\emph{A$_{1g}$}$(O$_{Tl-O}$) modes show a slight softening near $\emph{T}$$_c$ in Tl-2201, Tl-2212 and Tl-2223. Comparing to the other cuprate superconductors such as Bi$_2$Sr$_2$CaCu$_2$O$_{8+\delta}$$\cite{Boekholt}$, the $\emph{A$_{1g}$}$ mode at 650 cm$^{-1}$ showed an overall phonon hardening from 300 to 10 K. 

\section{DISCUSSION}

\subsection{Superconductivity-induced self energy effect}

In the conventional superconductor, the effect of superconductivity on the phonon modes is usually weak. The evolution of the  phonon frequencies in the microscopic picture is determined by some electronic excitations. Only the electronic states near the Fermi surface are changed as the sample enters the superconducting state. The self-energy effect$\cite{Zeyher}$ is due to the redistribution of the electronic density near the Fermi surface induced by the interaction between the phonon states and electrons. The effects are so small in cuprates that the phonons must be carefully measured with high accuracy and resolution\cite{Zhou, Boekholt, Burns}. According to the self-energy model$\cite{Zeyher, Nicol}$, the energy gap (2$\Delta$) opens due to superconductivity, producing the change in the self-energy $\Delta$$\Sigma$=$\Delta$$\omega$+$i$$\Delta$$\gamma$. The real part ($\Delta$$\omega$) represents the shift of the frequency, and the imaginary part ($\Delta$$\gamma$) is the variation in the linewidth. From the strong-coupling theory, the change of the phonon self-energy is affected by the ratio of $\omega$/2$\Delta$, and the corresponding results at different ratios are predicted. When $\omega$ $<$ 2$\Delta$, the phonon will be softened after entering the superconducting state, and the linewidth is expected to exhibit broadening. When $\omega$ is close to the position of 2$\Delta$, the frequency will show obvious softening near $\emph{T}$$_c$, and the linewidth can increase significantly. When $\omega$ $>$ 2$\Delta$, the frequency will show a monotonically increasing (hardening) trend, while the linewidth will continue to decrease.

It is necessary to determine the location of the superconducting gap before the phonon analysis. The bulk probes (Raman scattering, IR spectroscopy)$\cite{Devereaux, Wang}$ and the surface probes (electron tunneling spectroscopy, point-contact spectroscopy, break-junction tunneling spectroscopy, Andreev reflection spectroscopy)$\cite{Ozyuzer1, Ozyuzer2, Huang, Giubileo1, Giubileo2, Tsai, Takeuchi, Lanping}$ are the two major families of methods used to obtain 2$\Delta$ in the Tl-based materials. Tl-based compounds measured here are nearly optimally doped. Therefore, the 2$\Delta$ values should be used from the optimally doped Tl-based samples. Table \uppercase\expandafter{\romannumeral2} summarizes the early results from the tunneling spectroscopy\cite{Ozyuzer1, Ozyuzer2,Huang, Giubileo1, Giubileo2,Tsai, Takeuchi, Lanping}. 2$\Delta$ position increases with the increase of $\emph{T}$$_c$. For single-layer Tl-2201, $\emph{T}$$_c$ is between 85 and 91 K, $\Delta$ value obtained by point contact spectroscopy is about 19 to 25 meV, corresponding to 2$\Delta$ of about 306$-$403 cm$^{-1}$. The bilayer Tl-2212 has the $\Delta$ of about  20$-$28 meV, and the corresponding 2$\Delta$ is at 322$-$451 cm$^{-1}$ when $\emph{T}$$_c$ is in the range of 91$-$112 K. For tri-layer Tl-2223, the critical temperature is about 114$-$120 K, and the tunneling spectra give the $\Delta$ value at 23$-$35 meV,  yielding 2$\Delta$ in the range of 371$-$564 cm$^{-1}$.

The position of the $\emph{A$_{1g}$}$(Tl) mode for the studied Tl-2201, Tl-2212, and Tl-2223 is at 162, 133, and 132 cm$^{-1}$, respectively. Comparing to 2$\Delta$, which is generally higher than 300 cm$^{-1}$, the ratio of $\omega$/2$\Delta$ is less than 0.5. According to the self-energy model$\cite{Zeyher}$, the phonon mode in this range ($\omega$/2$\Delta$ = 0.2$-$0.5) will soften after entering the superconducting state, and the linewidth is expected to exhibit a little broadening. Referring to the analysis in Fig. 4, the self-energy model does not match the evolution of the observed behaviors of this low-frequency mode. Single-layer Tl-2201 and bilayer Tl-2212 exhibit phonon softening at $\emph{T}$$_c$. However, their linewidths stay unchanged at low temperatures. For tri-layer Tl-2223, the frequency is slightly hardened, and the corresponding linewith remains nearly constant below $\emph{T}$$_c$. Clearly, this $\emph{A$_{1g}$}$(Tl) mode does not obey the expected self-energy effect.

The  $\emph{A$_{1g}$}$(O$_{a}$) mode in Tl-2201, Tl-2212, and Tl-2223 is located at 491, 494, and 501 cm$^{-1}$, respectively.  Such values of $\omega$ are close to 2$\Delta$. It is thus expected to have a stronger self-energy effect. The ratio $\omega$/2$\Delta$ is about 1.2$-$1.6 in Tl-2201, thus the frequency should show a significant hardening near $\emph{T}$$_c$ according to self-energy model$\cite{Zeyher}$, accompanied by a significant broadening in the linewidth. From the  results on Tl-2201 as shown in Fig. 5, the frequency shows a smooth phonon softening at low temperatures, and the corresponding linewidth remains stable below $\emph{T}$$_c$. For bilayer Tl-2212, $\omega$/2$\Delta$ is 1.1$-$1.5. The frequency is expected to decrease, while the linewidth will show an obvious narrowing. The experiments give a decreasing frequency after entering the superconducting state and a nearly constant linewidth below $\emph{T}$$_c$. Finally, the $\omega$/2$\Delta$ of the trilayer sample is about 0.9-1.3. It is expected to have a significant softening in the frequency, accompanied by a broadening of the linewidth. From the temperature dependence of the Raman spectra (Fig. 5), the frequency of the tri-layer sample shows hardening near $\emph{T}$$_c$, while the linewidth shows a slight narrowing. Therefore, there is a big difference between the theoretical expectations from the self-energy model and experimental results.

\begin{figure}[tbp]
\includegraphics[width=\columnwidth]{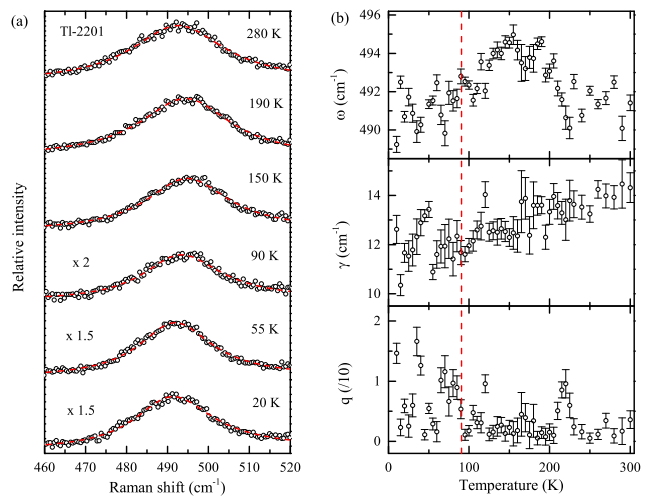}
\caption{(a) Representative Raman spectra (the black hollow circle) of Tl-2201 and the Fano fitting curve (red dash line) of the \emph{A$_{1g}$}(O$_{a}$) mode. (b) Temperature dependence of the $\omega$, $\gamma$, and $q$ value. The error bars correspond to  uncertainties in determining these parameters. The red vertical dashed line represents the $\emph{T}$$_c$ value.}
\end{figure}

The $\emph{A$_{1g}$}$(O$_{Tl-O}$) mode of Tl-2201, Tl-2212, and Tl-2223 is located at 603, 602, and 608 cm$^ {-1}$, respectively. These values for the high-frequency modes are much higher than the corresponding 2$\Delta$ positions. For single-layer Tl-2201, $\omega$/2$\Delta$ is about 1.5$-$1.97, the frequency is expected to show a monotonic increase on cooling, and the linewidth decreases. In Fig. 6, the frequency of Tl-2201 shows a slight softening at low temperatures, and the linewidth remains unchanged. For bilayer Tl-2212, $\omega$/2$\Delta$ is about 1.3$-$1.9, the frequency and linewidth tend to exhibit behaviors similar to those observed in the single-layer compound. However, the frequency of the bilayer sample shows a softening behavior in the superconducting state, and the linewidth in general remains unchanged. Since $\omega$/2$\Delta$ is about 1.1$-$1.6 for Tl-2223, the frequency is expected to decrease, while the linewidth will show an obvious narrowing. However, the results in Fig. 6 reveal that the trilayer sample still exhibits a smooth softening through $\emph{T}$$_c$, and the corresponding linewidth does not change significantly but rather keeping a nearly constant. Hence, the behaviors of this high-frequency mode do not follow the self-energy model as well.

All the selected modes in the studied Tl-based compounds do not follow the expected self-energy effect below $\emph{T}$$_c$. Instead, the universal phonon softening of the $\emph{A$_{1g}$}$(Tl), $\emph{A$_{1g}$}$(O$_{a}$), and $\emph{A$_{1g}$}$(O$_{Tl-O}$) mode at around 150 K is generally observed in all these Tl-based materials. The origin of this kind of phonon behavior might have important implications for the understanding of superconductivity in cuprates.

\begin{figure}[tbp]
\includegraphics[width=\columnwidth]{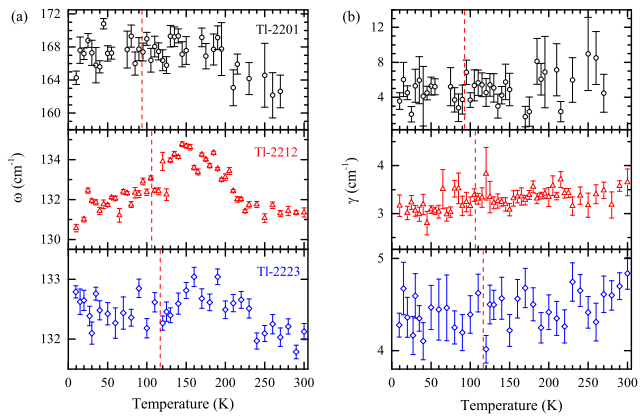}
\caption{Temperature dependence of the $\omega$ (a) and $\gamma$ (b) of the \emph{A$_{1g}$}(Tl) mode in Tl-2201 (black symbols), Tl-2212 (red symbols), and Tl-2223 (blue symbols). The red vertical dashed line represents the corresponding $\emph{T}$$_c$. }
\end{figure}

\subsection{Origin of the phonon softening}

The superconducting fluctuation may extend to the normal state above $\emph{T}$$_c$ in cuprate superconductors$\cite{Wen, Han}$. To prove the superconducting fluctuations above $\emph{T}$$_c$, one of the effective ways is to measure the Nernst effect$\cite{Xu}$. This method has been used to address the possible existence of vortex excitations in the pseudogap region of the cuprate superconductors. The theoretical work$\cite{Maki}$ based on the vortex liquid state suggests that the Nernst signal should exhibit an increase in the low-field region, and then it drops down when the magnetic field is close to a certain ratio to the upper critical field.

To find out the relationship between the phonon softening and the superconducting fluctuations, the onset temperature ($\emph{T}$$_{SF}$) for such fluctuations should be located. Some evidence for fluctuating superconductivity has been inferred from the measurements of the Nernst effect$\cite{Xu}$, in-plane angle-dependent resistance$\cite{LiuF}$, time-domain optical conductivity$\cite{Corson}$, and specific heat$\cite{Wen, Loram}$, etc. Table II gives the results of the measurements from various techniques$\cite{Duan1, Kim1, Duan2, Kim2, Wang2, Ma2}$. For single-layer sample Tl-2201, $\emph{T}$$_c$ was measured to be 87-90 K, and the corresponding $\emph{T}$$_{SF}$ is about 97.5 K. For bilayer Tl-2212, $\emph{T}$$_c$ of the prepared sample is 105$-$106 K, $\emph{T}$$_{SF}$ is near 120 K. For trilayer Tl-2223, $\emph{T}$$_c$ was measured to be 119 K and $\emph{T}$$_{SF}$ is about 135 K.

The universal phonon softening is observed at about 150 K from the temperature dependence of the Raman spectra. For single-layer Tl-2201, $\emph{T}$$_{SF}$ is much lower than this temperature. While for bilayer Tl-2212 and trilayer Tl-2223, $\emph{T}$$_{SF}$ is close to the initial softening temperature. Thus, phonon softening is not directly related to superconducting fluctuations for the single-layer compound. However, such a possibility can not be ruled out for the bilayer and trilayer samples.

The charge density wave (CDW) state appears in the normal state and was widely reported in cuprate superconductors$\cite{Tranquada, Hoffman, Wu, Ghiringhelli}$. Its origin is rooted in the instability of an one-dimensional system as described by Peierls. The extension of this concept to other systems has led to the concept of the Fermi surface nesting, which dictates the wave vector of the CDW and the corresponding lattice distortion$\cite{Zhu}$.

To detect the charge order in cuprate superconductors, resonant x-ray scattering (RXS)$\cite{Ghiringhelli,Comin,Silva}$, scanning-tunneling microscopy (STM)$\cite{Hoffman,Comin,Silva}$, and ARPES$\cite{Comin}$ were extensively carried out. However, atomic resolution STM images show that almost all the surface has a near-trigonal structure in optimally doped Tl-2212 and Tl-2223$\cite{Zhang}$. It is different from a checkerboard pattern in bilayer Bi-2212 single crystal$\cite{Hoffman,Silva}$. For single-layer Tl-2201, the CDW state was detected even in the overdoped region ($\emph{T}$$_c$=30 K). It was explained by a Fermi surface reconstruction, which is contrary to the case of under- and optimally-doped cuprates$\cite{Tam,Plat}$. Thus, the direct evidence for measuring the  CDW  state in optimally doped Tl-based materials is still lacking.

\begin{figure}[tbp]
\includegraphics[width=\columnwidth]{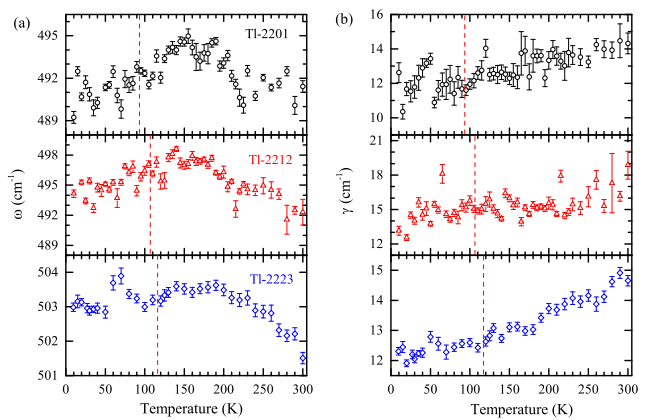}
\caption{Temperature dependence of the $\omega$ (a) and $\gamma$ (b) of the \emph{A$_{1g}$}(O$_{a}$) mode in Tl-2201 (black symbols), Tl-2212 (red symbols), and Tl-2223 (blue symbols). The vertical dashed line corresponds to $\emph{T}$$_c$.}
\end{figure}

Therefore, we have to focus on the pseudogap above $\emph{T}$$_c$. It is a consequence of two basic properties of high-$\emph{T}$$_c$ superconductors: The $d$-wave nature of the superconducting gap function varying as the cosine function around the Fermi surface, and the persistence of this gap into the normal state$\cite{Timusk}$.

ARPES$\cite{Ding, Harris,Kanigel}$, tunneling spectroscopy$\cite{Tao, Renner}$, nuclear magnetic resonance (NMR)$\cite{Kambe, Gerashenko, Zheng}$, and electrical transport$\cite{Takagi, Batlogg}$ can be directly used to detect the existence of the pseudogap in cuprates. Table II gives the onset temperature of the pseudogap ($\emph{T}$$^{PG}$) in the optimally doped Tl-based samples from NMR measurements\cite{Kambe,Gerashenko,Zheng}. The $\emph{T}$$_c$ value of the single-layer Tl-2201 is 85 K, while the $\emph{T}$$^{PG}$ value was determined at about  115$-$125 K\cite{Kambe}. For bilayer Tl-2212, $\emph{T}$$_c$ was measured to be 112 K, the corresponding $\emph{T}$$^{PG}$ was found around 130$-$170 K\cite{Gerashenko}. For trilayer Tl-2223, $\emph{T}$$_c$ of the as-grown sample is about 115 K, and the value of $\emph{T}$$^{PG}$ is about 140$-$160 K\cite{Zheng}.

\begin{table}[b]\scriptsize
\renewcommand\arraystretch{1.6}
\caption{Summary of the superconducting gap size (2$\Delta$), superconducting fluctuation temperature ($\emph{T}$$_{SF}$), and pseudogap onset temperature($\emph{T}$$^{PG}$) of Tl-based family from the works in the literature.}
\resizebox{84mm}{15mm}{
\begin{tabular}{cc c c c c c c c c c c c cccccccc}
\hline\hline
Sample & & &Tl-2201& & &Tl-2212& & & Tl-2223\\
\hline
$\emph{T}$$_c$(K) & & &85 - 91 & & &91 - 112& & &114 - 120\\
\hline
2$\Delta$(cm$^{-1}$) & & &306 - 403$\cite{Ozyuzer1, Ozyuzer2}$& & &322 - 451$\cite{Huang, Giubileo1, Giubileo2}$ & & &371 - 564$\cite{Tsai, Takeuchi, Lanping}$\\
\hline
$\emph{T}$$_{SF}$(K) & & &97.5$\cite{Duan1, Kim1}$ & & &120 - 150$\cite{Duan2, Kim2, Wang2}$& & &135 - 150$\cite{Ma2}$\\
\hline
$\emph{T}$$^{PG}$(K) & & &115 - 125$\cite{Kambe}$ & & &130 - 170$\cite{Gerashenko}$ & & &140 - 160$\cite{Zheng}$\\
\hline\hline
\end{tabular}}
\end{table}

For single-layer Tl-2201, $\emph{T}$$^{PG}$ is very close to the observed 150 K of the phonon softening temperature. Meanwhile, the temperature range of the pseudogap values covers the softening regime for Tl-2212 and Tl-2223. From this unexpected coincidence, it is not difficult to relate the observed softening to the interaction between the corresponding phonons and pseudogap. According to the band-structure calculations, pseudogaps are theoretically associated with the strong spin-phonon coupling$\cite{Jarlborg}$. In this model, the coulomb potential from the phonon modes and the spin-polarized part of the potential from the spin waves contribute to the opening of a partial gap at the Fermi energy. The phonon softening is used to estimate the obtained kinetic energy of the gap. This softening will be weak when the temperature is close to $\emph{T}$$_c$. Figures 4-6 show that the phonon softening does emerge in Tl-based family at a high temperature above $\emph{T}$$_c$ rather than entering the superconducting state. The theory also predicts that the atomic displacement will enhance the spin-phonon effect and the pseudogap value. The misalignment of atoms between the atomic layers in a normal state was indeed observed for the Tl-based samples$\cite{Dmowksi, Koyama, Koyama1}$. These vibrational displacements suggest a dramatic phonon softening far above $\emph{T}$$_c$. Besides, the specific heat measurements show that $\emph{T}$$_{SF}$ occurs between $\emph{T}$$_c$ and $\emph{T}$$^{PG}$$\cite{Wen, Loram}$ in the underdoped and optimally doped cuprate superconductors. This  is also consistent with the present results. That is, the phonon softening temperature is generally higher than $\emph{T}$$_{SF}$.

The obvious softening behaviors observed for Tl-based family are nicely consistent with the theoretical model within the strong spin-phonon coupling$\cite{Jarlborg}$. It is thus indicated that these universal softening trends are related to the pseudogap state from the analysis and comparisons of the existing experimental data.

\begin{figure}[tbp]
\includegraphics[width=\columnwidth]{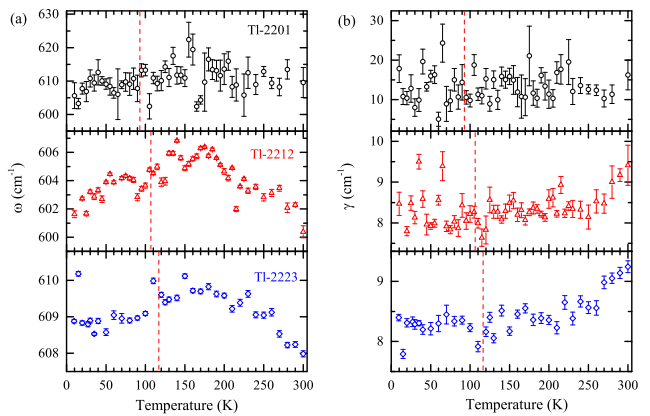}
\caption{Temperature dependence of the $\omega$ (a) and $\gamma$ (b) of the \emph{A$_{1g}$}(O$_{Tl-O}$) mode in Tl-2201 (black symbols), Tl-2212 (red symbols), and Tl-2223 (blue symbols). The vertical dashed line corresponds to $\emph{T}$$_c$.}
\end{figure}

\subsection{Possible phonon effect on the pseudogap}

The universal phonon softening far above $\emph{T}$$_c$ may provide evidence of the tight relationship between the superconductivity and the pseudogap. According to the promising theory for high-$T_c$ superconductivity\cite{Anderson}, pseudogap is a precursor pairing and originates only from the spin-charge coupling. In this case, there is only a superconducting pairing in the normal state and no phonon contribution in the pseudogap state is expected. $\omega$ and $\gamma$ are likely to exhibit the corresponding self-energy effect after entering the superconducting state. This picture seemingly gets the experimental supports for the feature of the magnetic order for the pseudogap\cite{zwang,fauq,yli}. However, the observed evolution of the phonon mode with temperature does not follow such the theoretical predictions. Along with the reported huge isotope effect on the opening temperature of the pseudogap\cite{Temprano, Bendele, Lanzara1}, the present results strongly indicate that the large lattice effect should be included in the understanding of the pseudogap phenomenon in cuprates. 

The anomalous softening behavior detected for the Tl-based family indicates the non-negligible lattice effect on the phonon modes. It seems that the opening of the pseudogap at least shares a part of the self-energy effect. As a result, the reduction of total self-energy and a weak self-energy effect when entering the superconducting state may follow$\cite{LimonovP}$. From our results, the decreasing frequency and unchangeable linewidth show a drastic self-energy loss at the temperature above $T_c$. Therefore, all the selected phonon modes, even the $\emph{A$_{1g}$}$(O$_{a}$) mode which is closest to 2$\Delta$, do not follow the expected behaviors in the superconducting state. In addition, the effect of the pseudogap is so powerful that it masks the superconductivity-induced phonon anomalies. It is why all the detected phonon modes continue to decrease during cooling when passing through the superconducting transition. All these features support the competition between the pseudogap and the superconductivity because they share in the same self-energy effect\cite{Hashimoto}.  Considering the evidence for the magnetic signature of the pseudogap\cite{zwang,fauq,yli,sugai2,Sugaimagnon,yli2,mans}, the present results might suggest the magnon-phonon coupling in the pseudogap state\cite{Struzhkin}. 

\section{CONCLUSIONS}

In conclusion, we have used Raman scattering spectroscopy to investigate the vibrational properties of high-quality Tl$_2$Ba$_2$Ca$_{n-1}$Cu$_n$O$_{2n+4+\delta}$ ($\emph{n}$$=$1,2,3) single crystals in the normal and superconducting state. The universal phonon softening at temperature near 150 K above the superconducting transition of each phonon mode has been generally observed. These phonon modes do not follow the expected self-energy effect. Instead, the feature can be related to the formation of the pseudogap state. The starting temperature for the phonon softening was found to be the opening temperature for the pseudogap. The results suggest the non-negligible lattice effect in the pseudogap state.

The work at HPSTAR was supported by the National Key R$\&$D Program of China (Grant No. 2018YFA0305900). The work at HIT was supported from the Shenzhen Science and Technology Program (Grant No. KQTD20200820113045081) and the Basic Research Program of Shenzhen (Grant No. JCYJ20200109112810241). Viktor Struzhkin acknowledges the financial support from Shanghai Science and Technology Committee, China (No. 22JC1410300) and Shanghai Key Laboratory of Material Frontiers Research in Extreme Environments, China (No. 22dz2260800).

\end{document}